\documentclass[twocolumn,prl,aps,superscriptaddress]{revtex4}
\usepackage{amssymb}

\usepackage{graphicx}

\usepackage{amsfonts}
\usepackage{amsmath}
\usepackage{epsfig}

\begin{document}

\title{Low-Energy Interference Structure with Attosecond Temporal Resolution}

\author{Xiaohong Song}
\affiliation{Research Center for Advanced Optics and Photoelectronics, Department of Physics, College of Science, Shantou
University, Shantou, Guangdong 515063, China}
\affiliation{Department of Mathematics, College of Science, Shantou University, Shantou, Guangdong 515063, China}
\affiliation{Key Laboratory of Intelligent Manufacturing Technology of MOE, Shantou University, Shantou, Guangdong 515063, China}

\author{Wenbin Jia}
\affiliation{Research Center for Advanced Optics and Photoelectronics, Department of Physics, College of Science, Shantou
University, Shantou, Guangdong 515063, China}

\author{Xiwang Liu}

\affiliation{Research Center for Advanced Optics and Photoelectronics, Department of Physics, College of Science, Shantou
University, Shantou, Guangdong 515063, China}
\affiliation{Department of Mathematics, College of Science, Shantou
University, Shantou, Guangdong 515063, China}

\author{Hongdan Zhang}

\affiliation{Research Center for Advanced Optics and Photoelectronics, Department of Physics, College of Science, Shantou
University, Shantou, Guangdong 515063, China}
\affiliation{Department of Mathematics, College of Science, Shantou
University, Shantou, Guangdong 515063, China}

\author{Qibing Xue}
\affiliation{Research Center for Advanced Optics and Photoelectronics, Department of Physics, College of Science, Shantou
University, Shantou, Guangdong 515063, China}

\author{Cheng Lin}
\affiliation{Research Center for Advanced Optics and Photoelectronics, Department of Physics, College of Science, Shantou
University, Shantou, Guangdong 515063, China}

\author{Wei Quan}
\email{charlywing@wipm.ac.cn}
\affiliation{State Key Laboratory of Magnetic Resonance and Atomic
and Molecular Physics, Wuhan Institute of Physics and Mathematics,
Chinese Academy of Sciences, Wuhan 430071, China}

\author{Jing Chen}
\email{chen_jing@iapcm.ac.cn}

\affiliation{HEDPS, Center for Applied Physics and Technology,
Collaborative Innovation Center of IFSA, Peking University,
Beijing 100084, China}

\affiliation{Institute of Applied Physics and Computational
Mathematics, P. O. Box 8009, Beijing 100088, China}

\author{XiaoJun Liu}
\email{xjliu@wipm.ac.cn}
\affiliation{State Key Laboratory of
Magnetic Resonance and Atomic and Molecular Physics, Wuhan
Institute of Physics and Mathematics, Chinese Academy of Sciences,
Wuhan 430071, China}

\author{Weifeng Yang}
\email{wfyang@stu.edu.cn}
\affiliation{Research Center for Advanced Optics and Photoelectronics, Department of Physics, College of Science, Shantou
University, Shantou, Guangdong 515063, China}
\affiliation{Department of Mathematics, College of Science, Shantou University, Shantou, Guangdong 515063, China}
\affiliation{Key Laboratory of Intelligent Manufacturing Technology of MOE, Shantou University, Shantou, Guangdong 515063, China}

\date{\today}

\begin{abstract}

Accessing precisely to the phase variation of electronic wave-packet
(EWP) provides unprecedented spatiotemporal information of
microworld. A radial interference pattern at near-zero energy has
been widely observed in experiments of strong-field photoionization.
However, the underlying physical picture of this interference
pattern is still under debate. Here we report an experimental and
theoretical investigation of this low-energy interference structure
(LEIS) in mid-infrared laser fields. We clarify that the LEIS arises
due to the soft-recollision mechanism, which was previously found
to play a pivotal role in producing the pronounced low-energy
structure. Specifically, the LEIS is induced by the
interference between direct and soft-recollision EWPs launched
within a 1/18 laser-cycle time scale in our experiments. Moreover,
the observation of LEIS is independent of laser wavelength and
specific atomic targets. Our result opens a promising new avenue for
retrieving the structure and dynamics of EWPs in atoms and molecules
with attosecond time resolution.

\end{abstract}


\maketitle

\section{Introduction}
Atomic photoionization under intense laser irradiation is a
fundamental process in strong-field light-matter interactions. With
the development of the electron-ion spectroscopy, especially in
momentum space, interferometric characterization of electron
wave-packet (EWP) has become possible. Interference effect of EWPs
has offered the potential for ultrafast imaging of atomic and
molecular structure and dynamics with unprecedented time resolution.
Above-threshold ionization (ATI) in which the electron reaches its
final continuum by absorbing more photons than the minimum required
for ionization \cite{Agostini1979}, exhibits the most prominent
strong-field interference effect. In this case, the interference
between two EWPs emitted within different laser cycles gives rise to
ATI peaks/rings spaced by one-photon energy in photoelectron
spectrum. Therefore, ATI can be taken as an intercycle interference
with the time resolution of one optical cycle. In another case, the
EWPs, released within one optical cycle, may also interfere with
each other and give rise to temporal double-slit interference
pattern, from which an subcycle time resolution can be
achieved \cite {Lindner2005,Gopal2009}. Recently, a holographic
structure has been observed in photoionization experiments at
mid-infrared wavelength and was attributed to the interference
between the direct and rescattered EWPs within the same
quarter-cycle of the laser pulse \cite{Huismans2011}.

Note that, for typical experiments that are implemented with
linearly polarized multicycle pulse, the intracycle interference
structures are usually entangled with the strong ATI fringes in the
photoelectron spectrum, and, as a consequence, are hardly recognized
or extracted in the experiments. Thus, the interference structures
in the very low energy region of photoelectron spectrum becomes in
particular interesting, as these low-energy structures have the
advantage of being well separated from the ATI fringes and are
potentially applicable for dynamic imaging. In this respect, it is
worth of mentioning that pronounced peaks in the laser polarization
direction, named as low-energy structure (LES), was recently
observed at low even near zero energy in the photoelectron spectrum
of atoms in strong infrared laser pulses
\cite{Blaga2009,Quan2009,Wu2012}. The LES was called as ``ionization
surprise'' \cite{Faisal2009} because it was beyond the scope of the
strong field approximation model \cite{keldysh,faisal73,reiss80} and
a closely related semiclassical picture
\cite{simpleman,Corkum1993,Schafer1993}, which have been well
adopted to understand various phenomena in strong-field physics.
Improved theoretical methods, e.g., classical trajectory Monte Carlo
(CTMC) model \cite{Quan2009,Liu2010,Wu2012,Kastner2012},
Coulomb-corrected strong-field approximation \cite{Yan2010}
and improved SFA with consideration of the Coulomb potential
\cite{Guo2013,WB14} have been applied to study the LES and revealed
that it can be attributed to Coulomb focusing effect
\cite{Quan2009,Wu2012,Liu2010} and bunching of photoelectron energy
caused by soft recollision \cite{Wu2012,Kastner2012,chen2002}.

Meanwhile, a radial interference structure in the low-energy region
of photoelectron momentum distribution (PEMD) has also been widely
observed in strong field ionization experiments. This structure was
implicitly shown in the early experiments of atoms (He, Ar and Ne)
\cite{Rudenko2004}. However, the origin of the structures remains
obscure so far
\cite{Arbo2006,Arbo2008,ZJChen2006,Gopal2009,Lai2017,Liu2012}. The
CTMC model attributes this interference pattern to the classical
angular momentum distribution and quantum analysis implies that it
is relevant to the minimum number of absorbed photons needed to
reach the threshold in multiphoton ionization \cite{Arbo2006,
Arbo2008,ZJChen2006}. Note that the electronic wave-packet(EWP) is
essentially a coherent matter wave and thus, its phase variation is
expected to play a significant role in the interference pattern.
Unfortunately, due to the classical essence of the above models, the
underlying physics related to the phase of EWP is beyond their
scope. In addition, CTMC model based on ADK tunneling theory is not
proper for description of the effect in multiphoton ionization
regime. In this respect, Lai \emph{et al}. applied a Coulomb
quantum-orbit SFA theory to study the low-energy fan-like structure
in 800 nm laser field and proposed a mechanism of subcycle
time-resolved holography, emphasizing the influence of the ionic
Coulomb potential on the phase of the forward-scattered electron
trajectories \cite{Lai2017}.

In this work, we experimentally and theoretically explore the
low-energy interference structure (LEIS) of atoms in mid-infrared
laser fields and aim to understand its underlying physical mechanism
comprehensively. This is accomplished by resorting to the
generalized quantum-trajectory Monte Carlo (GQTMC) simulations for
single ionization of xenon which agree well with the experimental
measurements for 1300 nm and 1800 nm laser fields. We identify that
the LEIS is induced by the interference between direct EWP and EWP
undergoing soft collision, which are released from a short time
window of 1/18 optical cycle around the maxima of the electric
field. This finding is of particular importance, considering the
fact that LEIS has some inborn advantages for investigating
ultrafast dynamics with attosecond time resolution. Since the energy
of the LEIS is far smaller than the photon energy, the interference
pattern will not be contaminated by its entanglement with the ATI
rings, as usually confronted by other intra-cycle interference
patterns, and can thus be easily extracted from the photoelectron
spectrum. In addition, LEIS is found to be a universal interference
effect, irrespective of specific atomic species and laser
parameters, such as wavelength and intensity. Therefore, LEIS can be
extensively used to image ultrafast dynamics of EWPs launched from
an extremely short temporal window with short-wavelength driven laser
pulse..

This paper is organized as follows. In Sec. II we introduce the
experimental setup and the GQTMC models. In Sec. III, we firstly
compare the experimental measurements and the GQTMC simulations on
LEIS. After that, with the semiclassical GQTMC statistical
trajectory-based analysis, we discuss the underlying mechanism of
the LEIS and the related intra- and intercycle interference pattern.
We summarize our results and conclude in Sec. IV.

\begin{figure*}
\centering
\includegraphics[width=0.8\textwidth,angle=0]{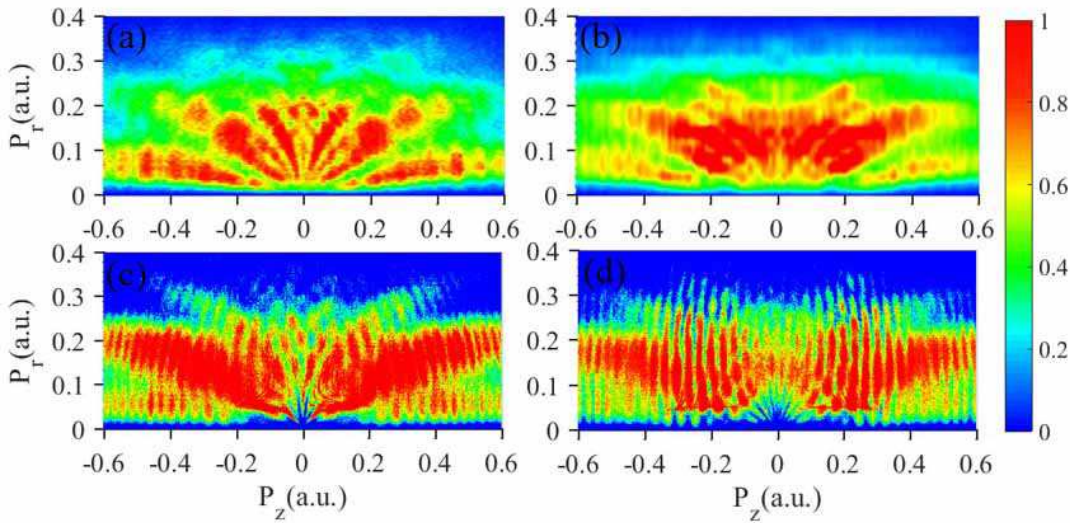}
\caption{ Comparison between theory and experiment in xenon (5p) ionized with intense ultrashort mid-infrared 30 fs laser pulse.
Comparison of experimental (top) and focal-averaged GQTMC theoretical (bottom ) results of PEMD obtained in xeon in similar conditions. (a) and (c): $\emph{I}=6.8 \times10^{13}$ $\mathrm{W/cm}^{2}$ and the wavelength $\lambda=1300$ nm. (b) and (d): $\emph{I}=7.2\times10^{13}$ $\mathrm{W/cm}^{2}$, $\lambda=1800$ nm.}
\end{figure*}

\section{Methods}

\hangafter=-1\hangindent=0pt\noindent\textbf{Experimental setup.}

The measurements have been performed with cold target recoil-ion
momentum spectroscopy (COLTRIMS) \cite{Ullrich2003,Jahnke2004}. A
commerical Ti: Sapphire femtosecond laser system (Legend,Coherent,
Inc.), which generates laser pulses with a center wavelength of 800
nm, a pulse duration of 35 fs and a repetition rate of 1 kHz, is
employed to pump a commerical OPA laser system (TOPAS-C, Light
Conversion, Inc.). The wavelength of the OPA laser system can be
tuned from 1100 nm to 2400 nm. The maximal output energies of the
laser pulses from OPA are about 1.0 mJ and 430 uJ for 1300 nm and
2400 nm, respectively. The pulse energy used in the ionization
experiments is controlled by means of an achromatic half waveplate
followed by a polarizer. The laser beam is directed and focused into
a supersonic Xe gas jet inside the COLTRIMS vacuum chamber.
Photoelectrons and Xe$^{+}$ ions are created in the laser focus and
extracted onto the time- and position-sensitive Microchannel plate detectors by a combination of a 2.3 V/cm electric field and a
2.3 Gauss magnetic field. The 3 dimensional vector momenta of
photoelectrons and Xe$^{+}$ were obtained from their times of flight
and impact positions.

\hangafter=-1\hangindent=0pt\noindent\textbf{Generalized quantum-trajectory Monte Carlo method.}

 With the development of ultrafast laser technology, highly nonlinear processes related to subcycle electron dynamics have been performed in experiments. Therefore, the ADK tunnel ionization rates using quasistatic approximation are no longer sufficient for exploring the experimental observation. Usually, the electron is assumed to be freed from the atom either via tunnel or multiphoton ionization. The electron escapes by climbing the top of the potential barrier along the energy axis \cite{Boge2013,Yu2005}, which is called as the 'vertical' ionization channel. In strong-field multiphoton ionization one is typically dealing with intermediate values of the Keldysh parameter $\gamma=1$ for most current intense field experimental conditions. Recent experiments have demonstrated the evidence of nonadiabatic tunneling at $\gamma\sim1$, and the dynamics are believe to deviate from adiabatic limit.

The GQTMC method is based on the nonadiabatic ionization theory \cite{Yudin2001,Perelomov1966}, classical dynamics with combined laser and Coulomb fields \cite{Brabec1996,Hu1997,Chen2000}, and the Feynman's path integral approach \cite{Salieres2001,MinLi2014}. The ionization rate is given as:

\begin{equation}
\Gamma(t)=N(t)\exp(-\frac{E_{0}^{\textbf{2}}f^{\textbf{2}}(t)}{\omega^{\textbf{3}}}\Phi(\gamma(t),\theta(t))).
\end{equation}
Here, $E_{0}f(t)$ and $\theta(t)$ are the envelope and the phase of the laser electric field, respectively. The preexponential factor is
\begin{eqnarray}
&N(t)=A_{n^{*},l^{*}}B_{l,|m|}(\frac{3\kappa}{\gamma^{\textbf{3}}})^{\frac{\textbf{1}}{\textbf{2}}}CI_{p}(\frac{2(2I_{p})^{\textbf{3/2}}}{E(t)})^{2n^{*}-|m|-1} \nonumber \\
\label{eq:12}\\
&\kappa=\ln(\gamma+\sqrt{\gamma^{\textbf{2}}+1})-\frac{\gamma}{\sqrt{\gamma^{\textbf{2}}+1}}\nonumber
\end{eqnarray}
where the coefficient $A_{n^{*},l^{*}}$ and $B_{l,|m|}$ coming from the radial and angular part of the wave function, are given
by Eq. (2) of Ref. \cite{Yudin2001}.
$C=(1+\gamma^{\textbf{2}})^{|m|/2+3/4}A_{m}(\omega,\gamma)$ is the Perelomov-Popov-Terent'ev correction to the quasistatic
limit $\gamma\ll1$ of the Coulomb preexponential factor with $A_{m}$ given by Eqs. (55) and (56) of Ref. \cite{Perelomov1966}. The
tunnelled electrons have a Gaussian distribution on the initial
transverse momentum
$\Omega(v_{r}^j,t_{0})\propto[v_{r}^j\sqrt{2I_{p}}/|E(t_{0})|]\exp[\sqrt{2I_{p}}(v_{r}^j)^{\textbf{2}}/|E(t_{0})|]$.
The coordinate of the tunnel exit shifts toward the atomic core, and $z_{0}=\frac{2I_{p}}{E(t_{0})}(1+\sqrt{1+\gamma^{\textbf{2}}(t_{0})})^{-\textbf{1}}$ is the tunnel exit point due to the nonadiabatic effects \cite{Perelomov1966}.
Thereafter, the classical motion of the electrons in the combined laser and Coulomb fields is governed by the Newtonian equation:
\begin{equation}
\frac{d^\textbf{2}}{dt^\textbf{2}}\mathbf{r}=-\mathbf{E}(t)-\bigtriangledown(\mathbf{V}(\mathbf{r})).
\end{equation}
Here, \textbf{V}(\textbf{r})  is the potential of the ion. According to the Feynman's path integral approach, the trajectory phase of the $\emph{j}$th electron is given by the classical action along the trajectory
\begin{equation}
{S_j} (\mathbf{p},t_0 )=\int_{t_0}^{+\infty}\{\mathbf{v_p^\textbf{2}}(\tau)/2+I_p-Z_{eff}/|\mathbf{r}(t)| \}d\tau
\end{equation}
where $\textbf{p}$ is the asymptotic momentum of the $\emph{j}$th electron and $Z_{eff}$ is the effective charge of the ionic core. The probability of each asymptotic momentum is determined by
\begin{equation}
{|\Psi|_\mathbf{p}}^\textbf{2}=|\sum_j\sqrt{\Gamma(t_0,v_{r}^j ) }\rm{exp}(\it{-iS_j})|^{\textbf{2}}.
\end{equation}
Using a parallel algorithm, one billion electron trajectories were calculated to obtain the PEMD. With the GQTMC method, we have reproduced some experimental and TDSE results \cite{Song2016,Yang2016,Lin2016,Song2017,Gong2017,Song2018}.

\section{Results and Discussion}

\begin{figure*}
\centering
\includegraphics[width=0.8\textwidth,angle=0]{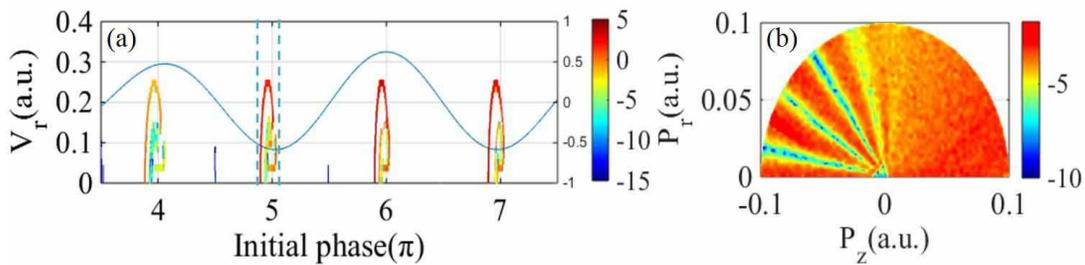}
\caption{(a): the initial tunneling coordinates of the photoelectrons which induces the LEIS ($|\textbf{p}|<0.1\ a.u.$)(b): The reconstruction of the finial momentum distribution with photoelectrons released from a single temporal windows. The conditions are similar to those used in Fig. 1(c).}
\end{figure*}

\hangafter=-1\hangindent=0pt\noindent\textbf{Low-energy interference structure.} Figures 1(a) and (b) show the typical photoelectron momentum distributions for xenon atoms measured by the COLTRIMS (see Methods) using intense ultrashort mid-infrared laser pulses (6.8 $\times10^{13}$ $\mathrm{W/cm}^{2}$, 1300 nm and 7.2 $\times10^{13}$ $\mathrm{W/cm}^{2}$, 1800 nm). We focus on the interference structure at low energy part ($<\hbar\omega$), where a radial fingerlike structure (dubbed LEIS in this work) near the threshold ($\textbf{p}_{z}=\textbf{p}_{r}=0$) can be identified. According to our measurements and documented works \cite{Rudenko2004,Arbo2006,Arbo2008,Gopal2009,Lai2017,Liu2012}, the LEIS is quite universal for experimental parameters and atomic species explored. In this study, to physically interpret the LEIS, theoretical simulations have been performed based on a full-dimensional GQTMC method with focal-averaging over the laser intensity (see Methods), where 3$\times10^{9}$ trajectories are launched. As shown in Figs. 1(c) and (d), the GQTMC simulations reproduce the measured LEIS surprisingly well, which lays a solid foundation to identify explicitly the origin of this LEIS.

\begin{figure*}
\centering
\includegraphics[width=0.8\textwidth,angle=0]{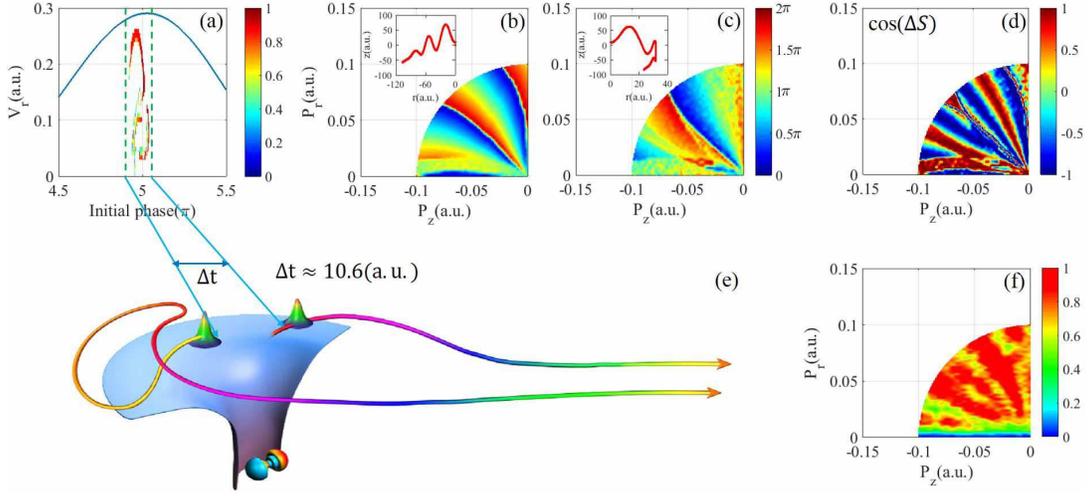}
\caption{ (a): the initial tunneling coordinates of the electrons $\textbf{p}_{z}<0$ in the LEIS region ($|\textbf{p}|<0.1\ a.u.$). (b)-(c): the phase of the direct and softly-recollided electron trajectories which are launched from the temporal window of (a), the insets of Fig (b) and (c) are the typical direct and rescattered electron trajectory, respectively.
(d): the phase difference between these direct and rescattered EWPs. (e): Artist's impression of LEIS. The direct and rescttered EWPs released from the temporal window ($<$ 10 a.u.) interfere with each other. (f): the experimental data.}
\end{figure*}


In order to shed light on the physical origin of the LEIS, the photoelectron trajectories have been traced with the semiclassical GQTMC back analysis of PEMD. For simplicity, the 2-dimensional photoelectron distribution of initial tunneling phase of laser field and transverse momentum for those producing the LEIS (i.e., $|\textbf{p}|<0.1\ a.u.$, where $\textbf{p}=\sqrt{p_{r}^{2}+p_{z}^{2}}$) at $6.8 \times10^{13}$ $\mathrm{W/cm}^{2}$ is presented in Fig. 2(a). As shown in this panel, photoelectrons contributing to the LEIS are launched from the sub-optical-cycle temporal windows around the peaks of the oscillating laser electric field. One of the temporal windows, indicated by two blue vertical dashed lines in Fig. 2(a), is chosen and the finial momentum distribution of the photoelectrons released from this window is pictured in Fig. 2(b). It is clear that the distinct interference fringes possessing the pattern close to those of LEIS can be identified for the momentum range of $\textbf{p}_z<0$. This result gives us a clue that the radial fingerlike structure might comes from the interference of the photoelectron trajectories launched from the identical sub-optical-cycle temporal window.

\begin{figure}[h]
\centering
\includegraphics[width=0.5\textwidth,angle=0]{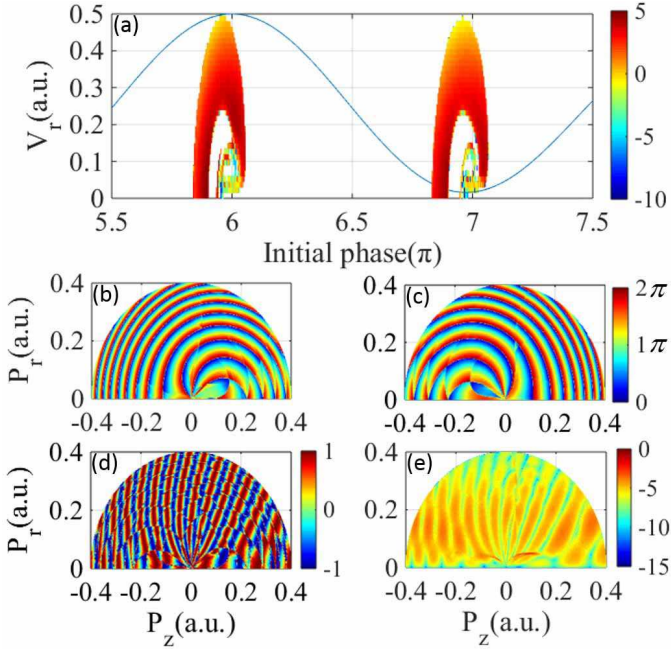}
\caption{ (a) the initial tunneling coordinates of the photoelectrons with energy ($0.4\ a.u.<|\textbf{p}|$ (b)-(c): the phases of EWPs from the two neighbor temporal windows. (d): the phase difference between these EWPs. (e): the reconstructed momentum distribution with
EWPs from the two temporal windows.}
\end{figure}

To comprehend how the interference of the photoelectron trajectories launched from the identical temporal window produces the LEIS, we further trace the initial tunneling coordinates of those electrons contributing to $\textbf{p}_z<0$ in the LEIS region ($0\leq\textbf{p}\leq0.1 a.u.$). As shown in Fig. 3(a), these photoelectrons come from a very short temporal window with duration of  $\sim$ 250 as (10.6 a.u.). According to the analysis, there are two types of electron trajectories: the first type is the direct electron trajectories where the photoelectron will be pulled away from the ion directly by the laser electric field (see the inset of Figs. 3(b)). As seen in the inset of Fig. 3(c), the other type of electron trajectories fulfills the condition of $z^{\star}\equiv{z(t^{\star})}\sim0$ and $p_z^{\star}\equiv{p_z(t^{\star})}=0$, where $t^{\star}, z^{\star}$ and $p_z^{\star}$ are the moment, position and momentum along laser polarization axis when the collision occurs. It is noteworthy that these electrons will experience soft recollision (see the inset of Fig. 3(c)) and energy bunching occurs, giving rise to the LES \cite{Kastner2012}.

\begin{figure*}
\centering
\includegraphics[width=0.8\textwidth,angle=0]{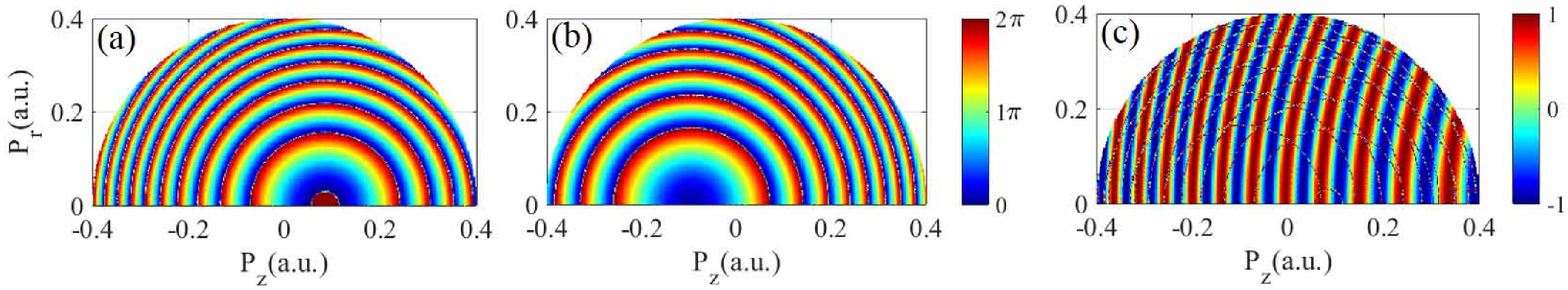}
\caption{(a) the phases of classical electron trajectories from two neighbor half laser cycles with $\mathbf{p}<0.4$ a.u. (b) The phase of electron trajectories presents concentric rings, the center of which occur in different place of the $\mathrm{p}_{\mathrm{z}}$ axis. (c) Phase difference between electron trajectories from two neighbor half cycles is achieved by $\cos \left[S_{1}(\mathbf{p})-S_{2}(\mathbf{p})\right]$ suggesting that these straight line fringes perpendicular to the laser polarization axis result from the interference between direct electron trajectories without considering the Coulomb potential in simple man model.}
\end{figure*}

The phases of these two types of quantum trajectories, which are obtained by the classical actions along their paths in the Feynman's path integral formulation, are presented in Figs. 3(b) and (c). Additionally, the radial fringes produced with the phase difference $\cos(\vartriangle S)$ are given in Fig. 3(d), which are coincident to the measurements (Fig. 3(f)) and numerical calculations (Fig. 2(d)). Thus, it can be understood that it is the interference between direct and rescattered EWPs launched from the same attosecond temporal window around the peak of laser electric field that induces the LEIS (see Fig. 3(e)). The fringes in the range of $\textbf{p}_z>0$ come from the same interference mechanism except that the laser electric field aligns in the opposite direction. This further explains the phenomenon in Ref. \cite{Gopal2009} that signatures of radial structures are only observed for $\textbf{p}_z<0$ with a carrier-envelope-phase (CEP) stabilized few-cycle laser pulse.

\hangafter=-1\hangindent=0pt\noindent\textbf{Interplay between the EWPs from different temporal windows.} In the following, we extend our GQTMC analysis to comprehend the interference of EWPs from different temporal windows. The initial conditions and phases of EWPs in two neighbor temporal windows for the photoelectrons in the range of $|\textbf{p}|<0.4$ a.u. are presented in Figs. 4(a), (b), and (c), respectively. The phase of EWPs emitted from each temporal window around the laser peak presents a concentric-ring-shape structure with a non-zero center on the $p_z$ axis and those of EWPs from two neighbour temporal window are symmetric to each other with respect to $p_{z}=0$, as shown in Figs. 4(b) and (c). On the other hand, the phase difference of the EWPs from two neighbor temporal windows is shown in Fig. 4(d), where straight lines almost perpendicular to the laser polarization axis can be identified. For comparison, the reconstructed photoelectron momentum distribution of EWPs released from the two temporal windows is given in Fig. 4(e), where the straight linear fringes close to those in Fig. 4(d) appear.

\begin{figure}[h]
\centering
\includegraphics[width=0.4\textwidth,angle=0]{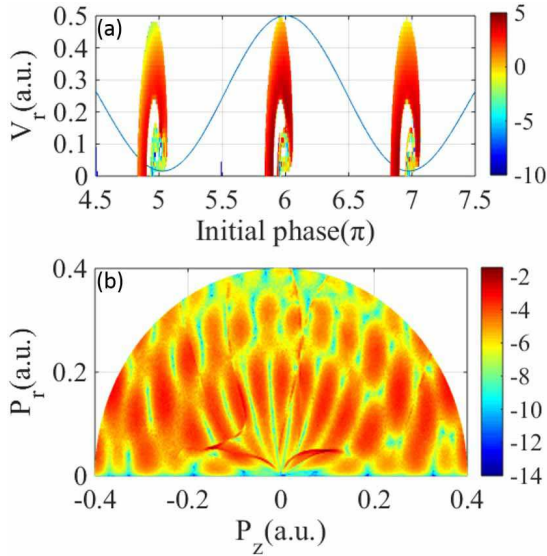}
\caption{(a) the initial tunneling coordinates of the photoelectrons from three neighbor temporal windows in $|\textbf{p}|<0.4$ a.u.(b) the reconstructed momentum distribution with the EWPs.}
\end{figure}

The calculations above indicate that these fringes can be attributed to the interference between direct electron trajectories of neighbour temporal windows. This is verified by the calculations with a simple man model \cite{Corkum1993}, where the Coulomb potential has been ignored. In this model, the classical actions \cite{Lewenstein1994} are
\begin{equation}
S_{j}(\mathbf{p},t_0 )=\int_{t_{0}}^{T_{p}}\left\{\frac{1}{2}[\mathbf{p}+\mathbf{A}(\tau)]^{2}+I_{p}\right\} d \tau ,
\end{equation}
where $\mathbf{p}=\mathbf{p}_{\mathbf{r}} \overrightarrow{\mathbf{e}}_{\mathbf{r}}+\mathbf{p}_{\mathbf{z}} \overrightarrow{\mathbf{e}}_{\mathbf{z}}$ denotes asymptotic momentum and the $t_{0}$ and $T_{p}$ are ionization moment and the end of the pulse. We can rewrite the classical action as
\begin{equation}
S_{j}(\mathbf{p},t_0)=\int_{t_{0}}^{T_{p}} \frac{1}{2}\left[\mathbf{p}_{\mathbf{z}}+\mathbf{A}(\tau)\right]^{2} d \tau+\left(\frac{1}{2} \mathbf{p}_{\mathbf{r}}^{2}+I_{p}\right) *\left(T_{p}-t_{0}\right)
\end{equation}.

In the simple man model \cite{Corkum1993}, the electron is born at the moment $t_{0}$ with the zero velocity along the direction of laser field, i.e.,$\mathbf{v}\left(t_{0}\right)=\mathbf{p}_{\mathbf{z}}+\mathbf{A}\left(t_{0}\right)=0$. From this equation we can obtain the relation between $t_{0}$ and $\mathrm{p}_{\mathrm{z}}$ numerically.

In Fig. 5 $(\mathrm{a})$ and $(\mathrm{b})$ , we show the phases of EWPs in two neighbor temporal windows calculated with the simple man model and the concentric rings close to the results in Fig. 4(b) and (c) can be found. The phase difference between two EWPs ionized from neighbor half cycles can be achieved by $\cos \left[S_{1}(\mathbf{p})-S_{2}(\mathbf{p})\right]$ \cite{Huismans2011}, as shown in Fig. 5$(\mathrm{c})$. There are fringes of straight lines almost perpendicular to the $\mathrm{p}_{\mathrm{z}}$ axis from $-0.4$ to $0.4$, which are very close to those in the reconstructed momentum distribution with EWPs coming from neighbor half cycles (see Fig. 4$(\mathrm{d})$) and also the previous experimental observation \cite{Gopal2009}, where the number of the relevant temporal windows has been chosen by compressing the pulse duration to 5 fs and stabilizing the carrier envelop phase.


To further demonstrate the validity of the interference mechanism discussed above, in Fig. 6(b), the momentum distribution for the EWPs from three neighbor temporal windows (Fig. 6(a)) has been calculated. As shown in this figure, the distinct ATI rings appear due to the intercycle interference can be identified, which is in good agreement with the measurements.

\section{Conclusion}
In summary, we investigate the low-energy interference structure in above-threshold ionization in mid-infrared laser fields both experimentally and theoretically. Our analysis clarifies that the LEIS arises due to interference between direct electron and electron experiencing soft recollision which are ionized within a small temporal window near the crest of the laser amplitude. The LEIS is universal in atoms and molecules and provides a potential way to retrieve structure and dynamics of EWP of atoms and molecules with attosecond time resolution.

\section{Acknowledgements}
The work was supported by the National Key Research and Development Program of China (No. 2019YFA0307700 and No. 2016YFA0401100), the NNSF of China (Grant No. 11674209, No. 11774215, No. 91950101, No. 11947243, No. 11774387, No. 11834015, and No. 11974383), Sino-German Mobility Programme (Grant No. M-0031), the Department of Education of Guangdong Province (Grant No.2018KCXTD011), the Open Fund of the State Key Laboratory of High Field Laser Physics (SIOM), and the Science and Technology Department of Hubei Province (No. 2019CFA035).

\end{document}